\documentclass[a4paper,11pt]{article}
\usepackage{pos}
\usepackage{braket}
\usepackage{graphics}
\usepackage{booktabs}
\usepackage{mathtools}
\usepackage{amsmath,amsthm,amsfonts,amscd} 
\usepackage{dsfont}
\usepackage{epstopdf}
\graphicspath{{./figs/}}
\usepackage{cleveref} 
\usepackage{wrapfig}
\usepackage{caption}
\usepackage{subcaption}
\newcommand{\mrgi}{M_{\text{RGI}}}
\newcommand{\msb}{\overline{\text{MS}}}
\newcommand{\amsb}{\alpha_{\msb}}
\newcommand{\mbar}{\overline{m}}
\newcommand{\Lrgi}{\Lambda_{\text{RGI}}}
\newcommand{\mom}{\mathcal{M}_n}
\newcommand{\dx}[2]{\text{d}^{#1}{#2}}
\newcommand{\tm}{\mu_{\mathrm{tm}}}
\title{A Quenched Exploration of Heavy Quark Moments and their Perturbative Expansion}
\ShortTitle{Quenched Heavy Quark Moments}

\author*[a,b]{Leonardo Chimirri}

\affiliation[a]{Humboldt Universit\"at zu Berlin, Institut f\"ur Physik \& IRIS Adlershof,\\ Zum Gro{\ss}en Windkanal 6, 12489 Berlin, Germany}

\affiliation[b]{DESY, Platanenallee 6, 15738 Zeuthen, Germany}

\emailAdd{leonardo.chimirri@desy.de}

\abstract{
	
	The parametric error on the QCD-coupling can be a dominant source of uncertainty in several important observables. One way to extract the coupling is to compare high order perturbative computations with lattice evaluated moments of heavy quark two-point functions. The truncation of the perturbative series is a sizable systematic uncertainty that needs to be under control.
	In this contribution we give an update on our study \cite{Chimirri:2022bsu} on this issue. We measure pseudo-scalar two-point functions in volumes of $L=2$ fm with twisted-mass Wilson fermions in the quenched approximation. We use full twist, the non-perturbative clover term and lattice spacings down to $a=0.010$ fm to tame the large discretization effects.
	Our results show that both the continuum extrapolations and the extrapolation of the $\Lambda$-parameter to the asymptotic perturbative region are very challenging.
}

\FullConference{%
  The 39th International Symposium on Lattice Field Theory (Lattice2022),\\
  8-13 August, 2022 \\
  Bonn, Germany 
}

\begin{document}
\maketitle
\section{The Strong Coupling}
Nearly all cross sections and decay rates of processes measured at LHC depend on some power of the strong coupling $\alpha(\mu)$, implying the presence of some parametric uncertainty on it. Prominent examples include the $Z$ boson total and partial widths \cite{Blondel:2018mad} and the $H\to gg$ and $H\to b\overline{b}$ partial widths \cite{Almeida:2013jfa}. In recent years lattice computations of the coupling reached an unprecedented precision and nowadays they dominate the world average \cite{DelDebbio:2021ryq}, requiring a meticulous assessment of all possible error sources. Here, we analyze quantitatively the size of uncertainties present when extracting $\alpha(\mu)$ from integrated, heavy quark two-point functions, a method first introduced by \cite{Bochkarev:1995ai,HPQCD:2008kxl}. This procedure relies on the simultaneous knowledge of this observable on the lattice and through high loop perturbative computations \cite{Maier:2009fz}, naturally prompting the question: within what energy window are both applicable with small and controlled errors? As a matter of fact, in a large physical box of length $L$, at a given number of simulation points $N=L/a$, the resolution of simulations is limited to wavelengths below the cutoff $\sim a^{-1}$. Typical values for $a$ result in scales at which the renormalized coupling's size may cause concern as to the size of the truncated part of the asymptotic series.

Herein, we study this issue in the quenched model -- arguing this to be qualitatively, but also quantitatively similar to the dynamical case -- carefully extrapolating to the continuum and finally estimating the truncation error of the perturbative expansion.

\section{Overview of Strategy} 
To start, let us define the central observable under scrutiny in continuum, Euclidean spacetime 
\begin{equation} \label{eq:mom-cont-def}
		\mom(\mrgi)=
		\int_{\mathds{R}}\dx{}{t}\,t^n\,G(t,m_h) =
		\int_{\mathds{R}}\dx{}{t}\,t^n\int\dx{3}{x}\,m_h^2\left\langle 
		P^{\dagger}(x)P(0) \right\rangle\,,\quad P(x)=i\,\overline{h}(x)\gamma_5h'(x)\,,
\end{equation}
which we refer to as \emph{n-th moment} -- an observable depending on one scale only. Here $m_h$ denotes the common bare mass of the heavy flavor doublet $(h,h')$ and if $Z_PZ_{m_h} = 1$ holds, then $J_{\text{RGI}}=\mrgi P_{\text{RGI}}=m_hP$. This, together with its small statistical noise, makes the pseudoscalar density a favorable, yet non-unique choice. To select what values of $n$ to study, note the following: (1) for pseudoscalar quantum numbers, one has $G(t,m_h)=G(-t,m_h)$ and odd $n$ moments vanish and (2) from power counting, or more precisely from a leading order OPE as $t\to\,0$, one obtains $G(t,m_h)\sim|t|^{-3}$ implying the $t$-integral to be finite for $n\ge4$. Altogether, one is left with values $n=4,\,6,\,8\,\ldots$ where we stress higher $n$ moments are dominated by longer distances, imposing an upper limit on $n$ if perturbation theory is to be employed.

The lattice moments' definition with a maximally twisted mass term, with periodic boundary conditions in space and open ones in time (at our small lattice spacings topological freezing may be worrisome \cite{Luscher:2011kk}) is
\begin{equation} \label{eq:mom-lat-def}
	\mom^{\mathrm{lat}}(a,\mrgi)=
	\lim_{L,T\to\infty}2\,a\sum_{t\in\mathcal{I}} t^n\,a^3\sum_{\vec{x}}
	\tm^2 \left\langle P^{\dagger}(t,\vec{x})P(0)\right\rangle\,,
\end{equation}
where $\tm$ is the bare twisted mass and the interval $\mathcal{I}$ consists of $t$-values far from the boundaries. Moreover, given the exponential decay of the correlator $G(t,m_h)$, noisy large $t$ contributions are suppressed and may be excluded. The absence of any dependence on the above choices has been extensively checked and confirmed within our precision. Finally, exploiting the above mentioned time-reversal symmetry one may sum over positive times only, with the factor 2 accounting for $t<0$.

Given the continuum limit of \eqref{eq:mom-lat-def}, it can be equated to the 4-loop perturbative expression \cite{Chetyrkin:1997mb,Maier:2009fz} of \eqref{eq:mom-cont-def} in the $\msb$-scheme
\begin{align}
	\label{eq:mom_expansion}
	&\lim_{a\to \,0}	\mom^{\mathrm{lat}}(a,\mrgi) \overset{\alpha\to 0}{\sim}
	\,\mbar^{\,4-n}_{\msb}(\mu) \sum_{i=0}^{3}c_n^{(i)}(s)\,\alpha^i_{\msb}(\mu)+
	\mathcal{O}\left(\alpha^4(\mu)\right)\,,\\
	\label{eq:mu_s}
	&\mu\equiv s\,\mbar_{\msb}(\mu)\,,
\end{align}
which for some fixed $s$ implies a one to one correspondence between $\mbar_{\msb}(\mu)$ and $\mrgi$. One can then invert \eqref{eq:mom_expansion} and \emph{extract} the coupling -- which we stress is an \emph{input parameter} of QCD -- up to an $\mathcal{O}(\alpha^4)$ truncation uncertainty of an a priori unknown size. This introduces a spurious $\mu$-dependence in the perturbative representation, which varies upon the variation of both the physical scale and the parameter $s$, yielding two possible handles to turn when trying to understand its behavior.\\

Let us define a constant physics trajectory by keeping 
\begin{equation}
	z\overset{\mathrm{def.}}{=}\sqrt{8t_0}\mrgi=
	\frac{\sqrt{8t_0}}{a}\frac{\mrgi}{\mbar_{\text{SF}}(\mu)}a\tm\left(Z^{\text{SF}}_P(a\mu,g_0)\right)^{-1} \left(1+\mathcal{O}\left(a\tm\right)\right)
\end{equation}
constant, where $\sqrt{8t_0}=0.463(3)$ fm is a gradient flow scale \cite{Luscher:2010iy} and renormalization factors at high $\beta$ are obtained from a reanalysis of the results of \cite{Capitani:1998mq}, carried out at half the original box length. The heavy quark mass $z$ sets the dominating scale of the observable and to inspect perturbation theory systematics we are interested in computing moments for various values of $z$, found in 
table \ref{table:masses}.
\begin{table}[htp!]
	\centering
	\caption{Approximate mass values in units of $\sqrt{8t_0}$ and of the quenched RGI-charm mass, given by $\mrgi^{\text{charm}}=1.684(60)\,\text{GeV}$ \cite{Rolf:2002gu}.}
	\label{table:masses}
	\begin{tabular}{c c c c c c}  
		\hline\toprule
		z& 13.5 & 9 & 6 & 4.5 & 3\\
    	\midrule
    	$\mrgi/\mrgi^{\text{charm}}$ &3.48 & 2.32 & 1.55&1.16 &0.77\\
		\bottomrule
\end{tabular}
\end{table}

Although automatic $\mathcal{O}(a)$-improvement is at play, large cutoff effects are expected and observed, prompting us to add a nonperturbatively \cite{Luscher:1996ug} tuned $c_{\text{SW}}(g_0)$ (which reduces higher order cutoff effects \cite{Frezzotti:2005gi,Becirevic:2006ii}) and to divide by the analytically\footnote
	{
		 This further modification is -- to the best of our knowledge -- common to all lattice results appearing so far in literature, although mostly computed numerically and not analytically.
	} 
computed tree-level at finite $a$ and $L$  
\begin{equation}
R_n(a\mrgi,\,z)\overset{\text{def.}}{=} 
\begin{cases}
	&\frac{\mathcal{M}_n(a\mrgi,z)}{\mathcal{M}_n^{\text{TL}}(a\tm^\text{TL},L/a)}\,,\quad n=4\\
	&\left(\frac{\mathcal{M}_n(a\mrgi,z)} {\mathcal{M}_n^{\text{TL}}(a\tm^\text{TL},L/a)}\right)^{\frac{1}{n-4}}\,,\quad n=6,\,8,\,10\,.
\end{cases}		
\end{equation}
Up to logarithmic corrections of the kind discussed in \cite{Husung:2019ytz}, this suppresses the leading $\mathcal{O}(a^2)$ effects in $G(t,m_h)$ by a factor $\alpha_{\text{bare}}$, but more care is needed for integrated observables \cite{Sommer:2022wac}. Some freedom is left in choosing $a\tm^{\text{TL}}$ and we previously \cite{Chimirri:2022bsu} observed $a\tm^\text{TL}=am_{\ast}$, where $m_{\ast}=\mbar_{\msb}(m_{\ast})$, to further improve the continuum approach. Note for $n>4$ the mass dimension $[R_n]=-1$ and beyond the fourth moment, further adimensional observables can be defined as \emph{ratios} of consecutive moments. Explicit mass prefactors present in \eqref{eq:mom_expansion} cancel in 
\begin{equation} \label{eq:dimless_moms}
\mathcal{R}_n(a\mrgi,z) =
	\begin{cases}
		&R_4(a\mrgi,z)\,,\quad n=4,\\
		&\frac{R_n(a\mrgi,z)}{R_{n+2}(a\mrgi,z)}\,,\quad n=6,\,8\,,
	\end{cases}
\end{equation}
leaving only a weak (logarithmic) dependence on the mass.

\section{Simulations} 
We briefly summarize here our setup and comment on the simulations. The pure Yang-Mills ensembles are generated with a plaquette action and consist of volumes of $L\sim 2$ fm, $T\sim6$ fm, while the topological charge is not integer valued with open boundary conditions. It has been monitored to check it actually varies freely and over a large enough range of values. To avoid disconnected diagrams we employ a mass degenerate doublet of Wilson-clover fermions with a twisted mass term, tuned to maximal twist by setting the Hopping parameter to its critical value. $\kappa_c$ is computed via fits of the data in \cite{Luscher:1996ug} and for each $\kappa(g_0)$ we tested independence on the fit function used and the number of points included in it, while including the known 1 loop coefficient, establishing the solidity of the procedure. The absence of boundary effects was monitored \cite{Chimirri:2022bsu} and table \ref{tab:gauge_runs} contains all important lattice parameters. W.r.t. to the status in \cite{Chimirri:2022bsu}, we also carried out dedicated studies at different $L$ for one of the coarser ensembles, establishing the absence of any visible finite volume corrections for quenched moments. 
\begin{table}[!htb] 
	\caption{Gauge run details, $l=L/a$, $t=T/a$, with plaquette definition for E(t).}
	\centering
	\label{tab:gauge_runs}
	\begin{tabular}{c c c c c c c c}
		\hline \toprule
		Run Name  &$\beta$&  $l^3\times t$& $N_{\text{cnfg}}$& $t_0/a^2$ &$a[fm]$ &$\tau_{\text{int}}(t_0)[\text{cfg}]$ & $\frac{\mathrm{GB}}{\mathrm{cnfg}}$\\
		\midrule[1.5pt]
		q\_beta616&6.1628 & $32^3\times96$  & 128  &  5.1604(98)& 0.071 & 0.78 & 1.7\\
		\midrule
		q\_beta628&6.2885 & $36^3\times108$ &  137 &  7.578(22) & 0.059 & 1.37 & 2.7\\
		\midrule
		q\_beta649&6.4956 & $48^3\times144$ &   109&  13.571(50)& 0.044 & 1.55 & 8.5\\
		\midrule[1.5pt]
		sft4      &6.7859 & $64^3\times192$ &  200 &  29.390(98) & 0.030 & 1.00 & 27 \\			
		\midrule
		sft5      &7.1146 & $96^3\times320$ & 80   &  67.74(23) & 0.020 & 0.55 & 152\\
		\midrule
		sft6      & 7.3600& $128^3\times320$&  98  & 124.21(91) & 0.015 & 1.03 & 360\\			
		\midrule
		sft7      & 7.700 & $192^3\times480$&  31  & 286.3(4.7) & 0.010 &-- & 1,823 \\			
		\bottomrule			
	\end{tabular} 
\end{table}

\section{Results}

\subsection{Continuum Limits} \label{subsec:contlims}
We will show here results for moments $\mathcal{R}_n$ defined in \eqref{eq:dimless_moms}, for $n=6,8$ and for the mass values in \cref{table:masses}. For $n=4$ a logarithmic enhancement of cutoff effects is present -- stemming from the integration over short distances -- for which another procedure was introduced as discussed at this conference in a separate contribution \cite{Sommer:2022wac} (note, this behavior is of interest also for $g-2$ and smoothed spectral function determinations).

\begin{figure} \label{fig:lowermoms} 
	\caption{Continuum limit of $R_6/R_8$, at high mass large cutoff effects may be present.}
	\begin{subfigure}{.5\textwidth}
		\centering
		\includegraphics[width=1.10\linewidth]{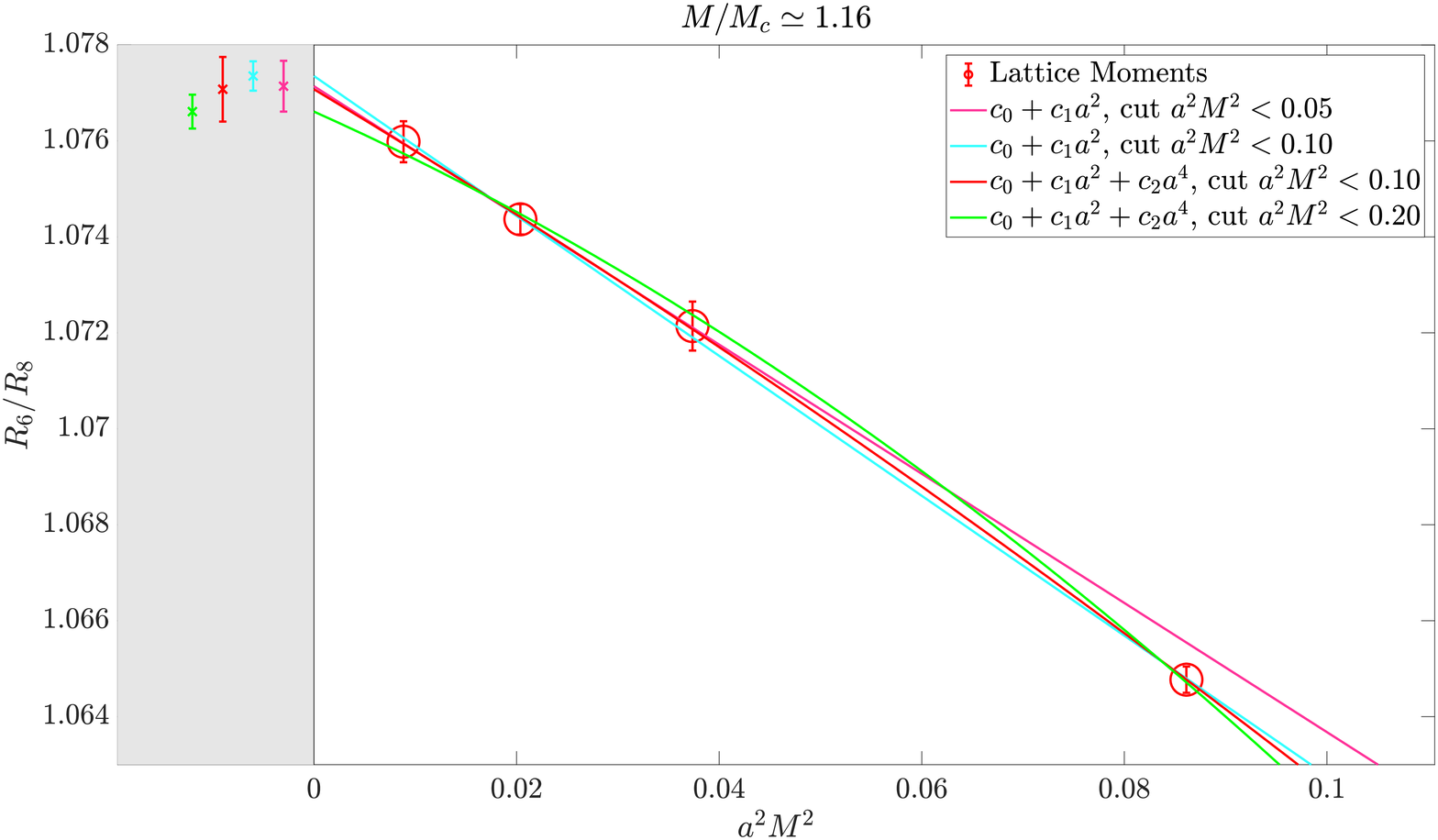} 
		\caption{Lower mass.}
		\label{fig:r6r8_z4p5_zoom}
	\end{subfigure}
	\begin{subfigure}{.5\textwidth}
		\centering
		\includegraphics[width=1.10\linewidth]{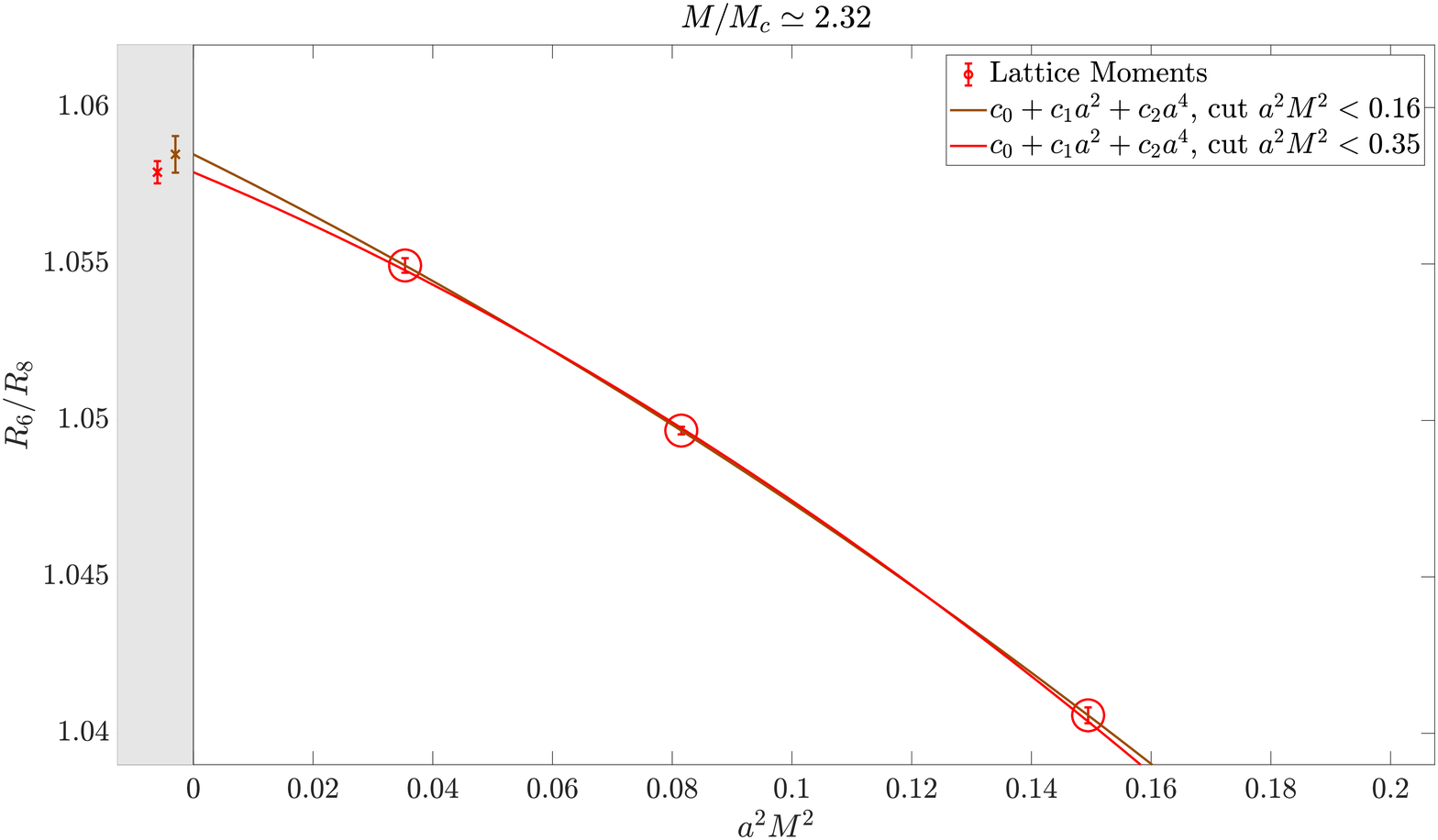} 
		\caption{Higher mass.}
		\label{fig:r6r8_z9_zoom}
	\end{subfigure}
\end{figure}

\begin{figure} \label{fig:highermoms}
	\caption{Continuum limit of $R_8/R_{10}$, cutoff effects here are under control.}
	\begin{subfigure}{.5\textwidth}
		\centering
		\includegraphics[width=1.10\linewidth]{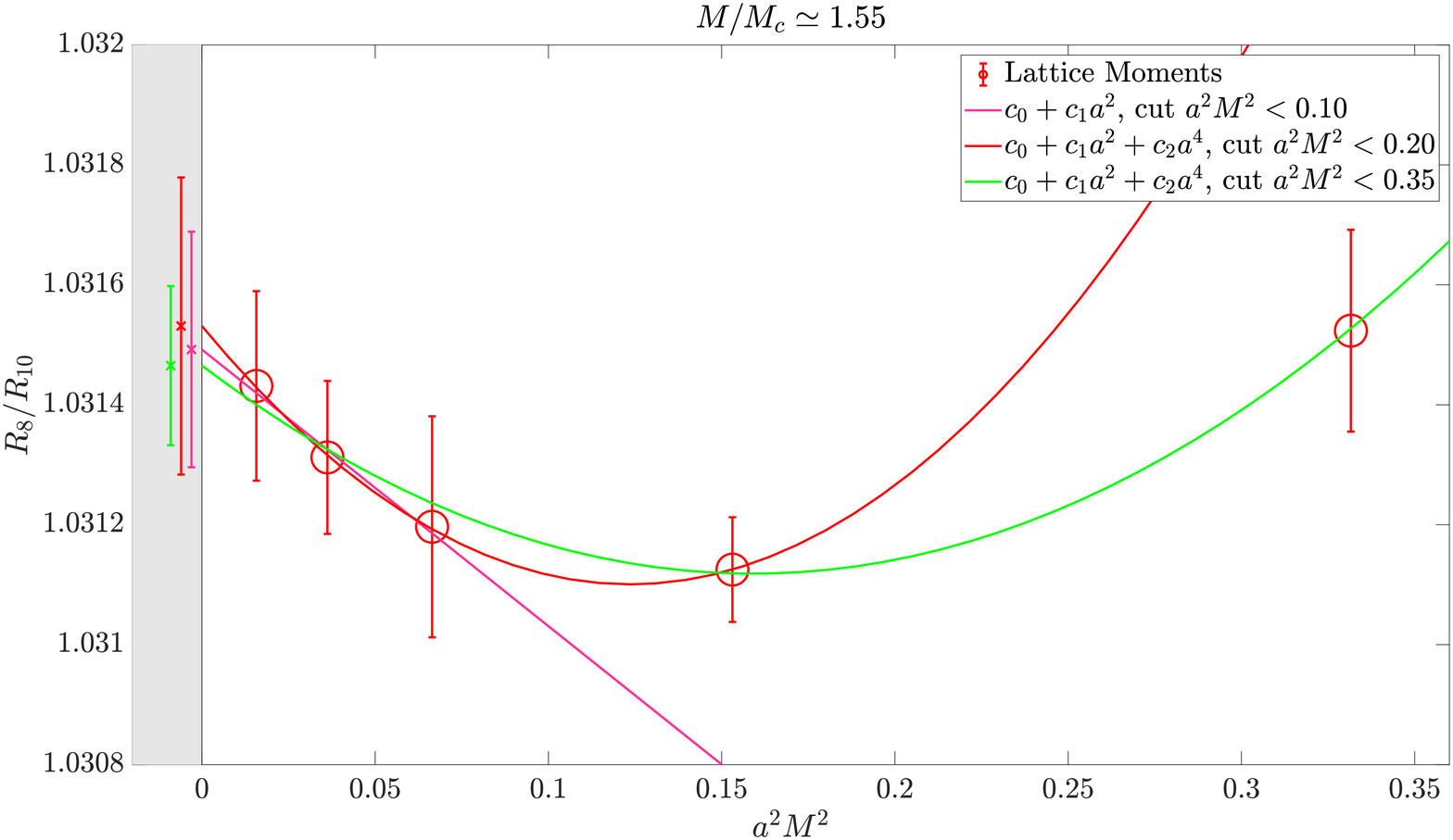} 
		\caption{Lower mass.}
		\label{fig:r8r10_z6_zoom}
	\end{subfigure}
	\begin{subfigure}{.5\textwidth}
		\centering
		\includegraphics[width=1.10\linewidth]{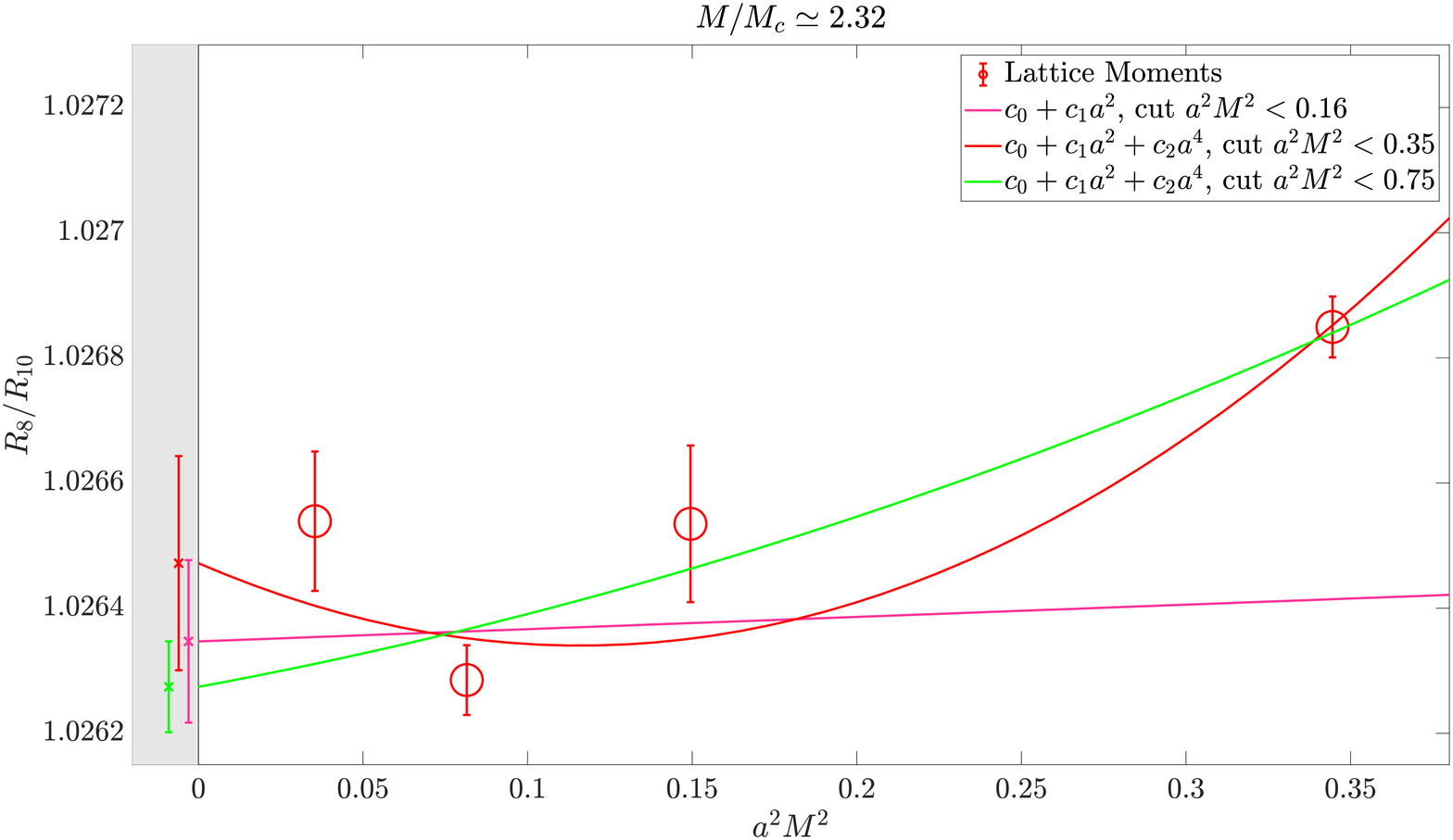} 
		\caption{Higher mass.}
		\label{fig:r8r10_z9_zoom}
	\end{subfigure}
\end{figure}

We fit linearly and quadratically in $(a\mrgi)^2$, for both Ans\"atze while varying the number of included points. We consider fits with a $p$-value of $p<0.05$ to have acceptable significance level and discard others. The continuum extrapolated values, depicted in \cref{fig:r6r8_z4p5_zoom,fig:r6r8_z9_zoom,fig:r8r10_z6_zoom,fig:r8r10_z9_zoom} in the gray band on the left hand side (next to each other for easier comparison), have a spread which is typically within the statistical error, yet sometimes quite an extrapolation is necessary from the point at smallest $a$ to $a\to\,0$. Take for instance \cref{fig:r6r8_z9_zoom}, 
where -- keeping in mind the sensitivity to the coupling is given by  $\mathcal{R}_n-1$, i.e. the distance from 1 -- one has an extrapolation of order 5\%. Knowing logarithmic enhancements of the type mentioned above to only be present at orders higher than $a^2$ for $n\ge6$ \cite{Sommer:2022wac}, still does not put us at ease in fully trusting the extrapolation; the 3 points closest to the continuum show (again, w.r.t. 1) relative cutoff effects of about $\sim30\%$. We thus decide to give quite a conservative estimate, namely by taking half of the extrapolated distance in the y-axis and adding it in quadrature to the statistical error. Let us mention for $z=13.5$, the highest mass, we were able to extrapolate for $R_8/R_{10}$, but not for $R_6/R_8$.
\subsection{Extraction of the Coupling} \label{subsec:extract_coupl}
Our goal is to study the applicability domain of perturbation theory to moments, for which we focus here on the coupling. Adimensional observables may be expanded as 
\begin{equation}
	\mathcal{R}_n(\mrgi)\overset{\alpha\to\,0}{\sim}1+
	d_n^{(1)}\amsb(\mu) +
	d_n^{(2)}(s)\,\amsb^2(\mu) +
	d_n^{(3)}(s)\,\amsb^3(\mu) + \mathcal{O}(\amsb^4(\mu))\,,\quad
	\mu\equiv s\,\mbar_{\msb}(\mu)\,,
\end{equation}
and inverting for each $s$, several values for $\amsb(\mu)$ can be extracted, \emph{all with an intrinsic} $\mathcal{O}(\alpha^4)$ \emph{uncertainty}. At a given $s$, the size of the truncated part is determined by two factors: the unknown coefficient $d_n^{(4)}(s)$ (plus higher order corrections) and by the coupling's value at a given scale. How the size of the truncated part \emph{changes with the scale} is, in turn, encoded in the $\beta_{\msb}(\amsb)$-function; quantitatively put, in the asymptotic region
\begin{align}
	&\beta_{\msb}(\amsb) \overset{\alpha\to\,0}{\sim}
	-\beta_0\,\amsb^{\,2}(\mu)-\beta_1\,\amsb^{\,3}(\mu) + \mathcal{O}(\alpha^{\,4})\,,\quad\text{with}\\
	&\big[\beta_0,\,\beta_1\big]= 
	\left[\frac{1}{(4\pi)}\left(11-\frac{2}{3}N_f\right),\,
	\frac{1}{(4\pi)^2}\left(102-\frac{38}{3}N_f\right)\right] \simeq 
	\begin{cases}
		\big[0.88,\, 0.65\big]\,,\,\, N_f=0\\ 
		\big[0.66,\, 0.33\big]\,,\,\, N_f=4
	\end{cases} \,,
\end{align}
so that there is strong perturbative indication to expect a very similar behavior, qualitatively but also quantitatively, between quenched and the fully dynamical (lattice accessible) case, namely $N_f=4$. To conclude, a quenched computation is a well motivated first step into studying the size of the asymptotic scaling region of moments.

\subsection{$\Lrgi$'s Asymptotic Scaling}
\begin{figure} \label{fig:lambdafits} 
	\caption{Results of scaling violations in $\alpha$ in the $\Lambda$ parameter for the two main observables, built with appropriate ratios of lattice-normalized moments. Description in the main text.}
	\begin{subfigure}{.5\textwidth}
		\centering
		\includegraphics[width=1.03\linewidth]{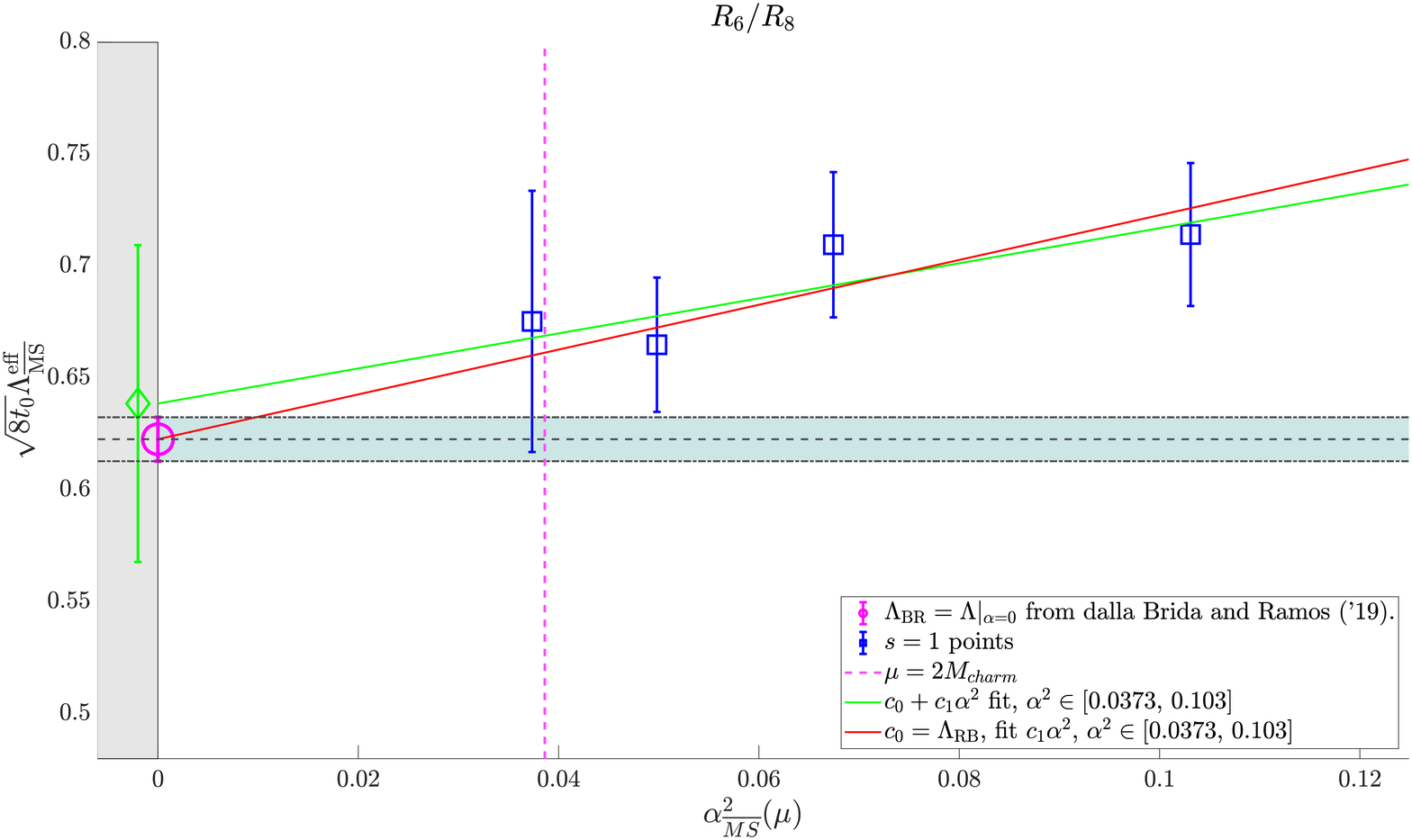} 
		\caption{Fit for $\Lambda_{\msb}$ as a function of $\amsb^2$, compared with\\ a fit constrained to pass through $\Lambda_{\mathrm{BR}}$ in \cite{DallaBrida:2019wur}, extracted\\ from $R_6/R_{8}$.}
		\label{fig:fit_r6r8extra}
	\end{subfigure}
	\begin{subfigure}{.5\textwidth}
		\centering
		\includegraphics[trim=74 0 00 00, clip, width=0.97\linewidth]{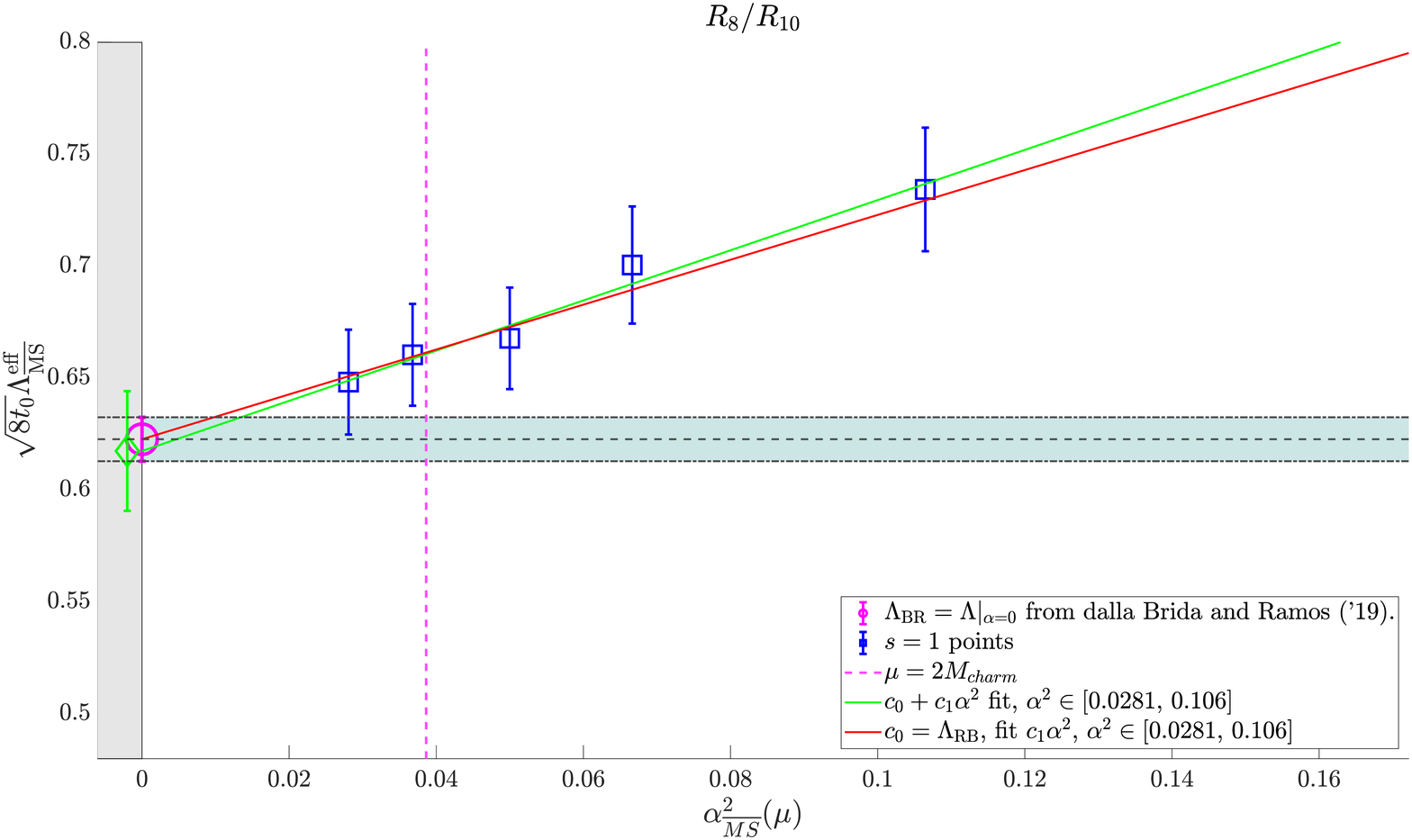} 
		\caption{Fit for $\Lambda_{\msb}$ as a function of $\amsb^2$, compared with a fit constrained to pass through $\Lambda_{\mathrm{BR}}$ in \cite{DallaBrida:2019wur}, extracted from $R_8/R_{10}$.}
		\label{fig:fit_r8r10extra}
	\end{subfigure}
\end{figure}

Here, instead of results for $\mathcal{R}_n$ as a function of $\alpha^4$, we directly show results for the $\Lrgi$-parameter, computed through\footnote{Here we prefer the usage of $g^2=4\pi\alpha$.} its ratio with $z=\sqrt{8t_0}\mrgi$ -- which we remind here was a chosen value we tuned our bare masses to -- as
\begin{align} 
	\sqrt{8t_0}\Lambda_{\msb}\,=\, 
	&\,z\cdot s\cdot\frac{\left(b_0 g^2_{\msb}(\mu)\right)^{-b_1/(2b_0^2)}} {\left(2b_0g^2_{\msb}(\mu)\right)^{-d_0/(2b_0)}}
	\exp\left\{-\frac{1}{2b_0g^2_{\msb}(\mu)}\right\} \cdot\\
	&\cdot\exp\left\{ -\int_{0}^{g_{\msb}(\mu)} \text{d} x \left[\frac{1-\tau_{\msb}(x)}{\beta_{\msb}(x)}+ \frac{1}{b_0x^3} -\frac{b_1}{b_0^2x}+\frac{d_0}{b_0x} \right] \right\}\,,
\end{align}
with the 5 loop $\beta_{\msb}(g)$ \cite{Herzog:2017ohr,Luthe:2017ttg} and 4 loop quark mass anomalous dimension $\tau_{\msb}(g)$ \cite{Chetyrkin:1997dh,Vermaseren:1997fq}, and where we plug in the value of $g_{\msb}$ obtained in \cref{subsec:extract_coupl}. The leftover $\alpha^4$ uncertainty in $\mathcal{R}_n$ implies, at leading order, an asymptotic scaling of $\Lrgi$ as
\begin{align}
	\Delta\Lambda=\frac{\text{d}}{\text{d} \amsb} \left(\frac{\Lambda}{\mu}\right) \Delta(\amsb)
	 = \frac{c}{\amsb^2} \amsb^4\left(1+\mathcal{O}(\amsb)\right)\,\implies\,\Lambda_{\msb}^{\text{eff}}\,=\,
	\Lambda_{\msb}+\mathcal{O}\left(\alpha^2_{\msb}\right)\,,
	\label{e:Lameff}
\end{align}
where $c$ is some constant. We thus show $\Lambda_{\text{eff}}$ vs. $\amsb^2(\mu)$, both obtained from the moments' ratios under scrutiny, in \cref{fig:fit_r6r8extra,fig:fit_r8r10extra}. We show only results for $s=1$, where, to stay on the safe side, we select the extracted $\amsb$ values with the largest errors. The purple value is the result of \cite{DallaBrida:2019wur}, which we indicate with $\Lambda_{\mathrm{BR}}$, and it is valid for $\alpha=0$, but we add a light blue horizontal line to guide the eye. We show two different fits, a free fit linear in $\alpha^2$ in light green -- with the extrapolated result drawn in the light gray band -- and, again, a fit linear in $\alpha^2$, but constrained to pass through $\Lambda_{\mathrm{BR}}$, in red. Finally, a dashed vertical light purple line indicates the scale $\mu\simeq2m_{\text{charm}}$.\\ 
First, we start by noticing the 
uncated part to be quite large at a considerable energy scale. As expected, higher moments have a larger deviation w.r.t. the asymptotic $\alpha\to\,0$ value, but they can rely on more precise lattice results -- this is nothing more than the statement of the window problem we are dealing with. For the free fit, we obtain results completely compatible with \cite{DallaBrida:2019wur} (but also with other pure gauge studies, such as \cite{Asakawa:2015vta,Kitazawa:2016dsl,Ishikawa:2017xam}), but not with a particularly competitive error. The constrained fit agrees very well with the unconstrained one, validating the theoretically motivated Ansatz of $\alpha^2$ scaling and giving supporting evidence of overall consistency.
\section{Conclusions}
We have computed the strong coupling from moments of heavy-quark two point functions over a range of energies from slightly below the charm mass up to $\mu\sim3.5\,m_{\text{charm}}$. The continuum limit was difficult and lattice spacings down to $a\sim0.01$ fm were used (and in one case, for $n=4$, a further approach had to be developed \cite{Sommer:2022wac}), enabling us to extrapolate reliably in almost all cases.

The objective was to study quantitatively the behavior of unknown, higher order corrections in the coupling, which are truncated after the known term of order $\alpha^3$ \cite{Maier:2009fz}, to compute around what energy scale these are small or close to absent within errors. As a matter of fact, studies of cases where $\alpha$ needs to be unexpectedly small for this to happen exist \cite{DallaBrida:2016uha} and are part of our overall motivation.

At the scale $\mu\simeq2m_{\text{charm}}$, the deviations are worryingly of order $\sim10\%$, where we stress once more these results are obtained in the quenched theory and cannot currently be reproduced with comparable precision in the fully dynamical case.

Differently put, turning the reasoning around, the moments method to extract the strong coupling and the charm or bottom quark mass would greatly benefit in precision and reliability if its 5 loop computation was available.  
\paragraph{Acknowledgements}
This project has received funding from the European Union’s Horizon 2020 research and
innovation programme under the Marie Skłodowska-Curie grant agreement No. 813942.
\bibliographystyle{./JHEP}
\bibliography{./refs}
\end{document}